# Inverse Lithography Physics-informed Deep Neural Level Set for Mask Optimization

Xing-Yu Ma,[1] Shaogang Hao,[1,*]

[1]*Tencent Quantum Lab, Tencent, Shenzhen, Guangdong 518057, China.*

**shaoganghao@tencent.com*

As the feature size of integrated circuits continues to decrease, optical proximity correction (OPC) has emerged as a crucial resolution enhancement technology for ensuring high printability in the lithography process. Recently, level set-based inverse lithography technology (ILT) has drawn considerable attention as a promising OPC solution, showcasing its powerful pattern fidelity, especially in advanced process. However, massive computational time consumption of ILT limits its applicability to mainly correcting partial layers and hotspot regions. Deep learning (DL) methods have shown great potential in accelerating ILT. However, lack of domain knowledge of inverse lithography limits the ability of DL-based algorithms in process window (PW) enhancement and etc. In this paper, we propose an inverse lithography physics-informed deep neural level set (ILDLS) approach for mask optimization. This approach utilizes level set based-ILT as a layer within the DL framework and iteratively conducts mask prediction and correction to significantly enhance printability and PW in comparison with results from pure DL and ILT. With this approach, computation time is reduced by a few orders of magnitude versus ILT. By gearing up DL with knowledge of inverse lithography physics, ILDLS provides a new and efficient mask optimization solution.

# 1. INTRODUCTION

As the feature size of integrated circuits (IC) continues to shrink, optical proximity effects in the lithography process have become increasingly significant, greatly impacting IC fabrication yields. Computational lithography physical modeling and resolution enhancement technology (RET) [1] have emerged as crucial tools for enhancing printed pattern fidelity and process window (PW), ensuring complex IC pattern printing and yield, particularly in advanced process. RET methods such as sub-resolution assist features [2], phase-shifted masks [3], off-axis illumination [4] and optical proximity correction (OPC) [5], are employed to optimize mask for precise wafer image.

As the most widely used RET, OPC utilizes empirical or numerical algorithms to compensate pattern errors caused by optical interference, diffraction effects, three-dimensional thick mask effects, light polarization effects, projection lens aberrations, and lithography process variations (e.g., depth of focus and exposure dose variations) by correcting mask geometrical features. [6]. OPC could be broadly categorized into the three types: rule-based OPC method [7], model-based OPC method [8] and inverse lithography technology (ILT) [9-11]. Rule-based OPC method involves firstly determining correction rules based on experimental and simulation data as well as engineering experience, then correcting local mask geometry features such as edges and corners according to these rules. Although rule-based OPC method has been successful for years, the correction variables are too simplistic to meet lithography process requirements for process nodes of 90 nm and below. Model-based OPC method relies on lithography imaging models and optimization algorithms to adjust edge positions of mask pattern, achieving higher mask printability than rule-based OPC method. To further improve print fidelity of mask for smaller process nodes, it is necessary to refine mask modifications with greater precision. ILT pixelates the mask pattern and optimizes the pixel values of mask through optimization methods like conjugate gradient descent (CGD) method, further enhancing mask fidelity. ILT has been employed to optimize extreme ultraviolet lithography masks, significantly improving overall PW [12,13]. ILT is divided into two categories: pixel-level explicit method [14,15] and level set implicit methods [10,11]. The former optimizes each mask pixel and results in a large number of tiny isolated and edge-slit structures in the mask, leading to low manufacturability in high volumes due to unmanageable mask writing times. In contrast, the latter evolves the mask boundary and usually produces relatively clean masks and further improves manufacturability through Manhattanized mask optimization. On the other hand, as ILT is based on the optimization of numerous mask pixel variables, it requires a lot more computation resource and run time. Therefore, it is a great challenge to directly apply ILT to full-chip mask optimization, and it is currently primarily used for fine-tuning specific layers and hotspot regions.

Along with the rapid development of machine learning (ML) technology in recent years, IC industries have been actively adopting ML methods in computational lithography modules, such as lithography imaging model modeling [16], hotspot detection [17], SRAF [18], pattern selection [19], OPC [20], and notably, the acceleration of ILT. Variety of ML methods have been applied to different problems, for examples, the first application of conditional generative adversarial networks for pixel-level mask optimization [21], pixel-level mask optimization utilizing graphical neural network [22], the Neural-ILT method that combines deep learning (DL) with pixel-level ILT for mask optimization [23], the DevelSet method that employs DL model for level set-based ILT [24], the model-driven DL method for pixel-level mask optimization [25] and etc. Among different ML methods, DL holds a lot of attention and has been introduced to pixel-level ILT. However, the masks generated by DL aided pixel-level ILT are in many cases too complex and difficult to manufacture and integrate into the commercially viable level set-based ILT frameworks. Additionally, this method often generates initial masks with low pattern fidelity, affecting further fine optimization and PW enhancement.

In this paper, we propose an inverse lithography physics-informed deep neural level set method (ILDLS) for mask optimization. We design the level set-based ILT correction layer as a layer of DL integrated into the DL generation mask framework. During forward propagation of the model, the fidelity error can be calculated based on wafer pattern generated from the lithography imaging model. During back propagation, the gradient of the level set-based ILT layer updates the neuron weights through the chain rule, enabling the neuron weights of the model to implicitly contain the domain knowledge of inverse lithography physics, thus improving the pattern fidelity and PW. Simulation results demonstrate that this method substantially improves pattern fidelity and PW compared to pure DL and level set-based ILT. The computational efficiency of level set-based ILT is improved by two to three orders of magnitude. This method seamlessly combines DL and level set-based ILT with domain knowledge from inverse lithography physics, providing a novel level set-based ILT acceleration strategy that has attracted widespread interest in bridging the new solution to commercial level set-based ILT frameworks.

The remainder of the paper is organized as follows: Section 2 introduces the lithography simulation model and the level set method. The proposed method is described in Section 3. Results and discussion are presented in Section 4. Conclusion is provided in Section 5.

## 2. PRELIMINARIES

In this section, we present an overview of the lithography simulation model and the level set approach.

### A. Lithography Simulation Model

The lithography simulation model is composed of a lithography imaging model and a photoresist model. The lithography imaging model simulates the distribution of light intensity on the wafer, also known as the aerial image. In our study, we employ the Hopkins optical model [26] to represent a partially coherent imaging system (a 193 nm wavelength system in this paper), which is widely adopted due to its low computational cost and high reliability. Theoretically, the optical model can be decomposed into a sum of coherent optical systems through singular value decomposition (SVD). The optical projection process can be represented by a series of coherent optical kernels, allowing the aerial image to be obtained by convolving a mask with these coherent optical kernels, as expressed in the following equation:

$$I(x, y; h_\mu) = \sum_{k=1}^{K} \omega_k(h_\mu)|M(x,y) \otimes h_k(x, y; h_\mu)|^2 \quad (1)$$

where $h_k(x, y; h_\mu)$ and $\omega_k(h_\mu)$ are optical kernels and the corresponding weights obtained by SVD respectively, $h_\mu$ is the defocus, the $M(x, y)$ and $I(x, y; h_\mu)$ are the mask and aerial image, respectively, and $\otimes$ represents the convolution operation. The partially coherent optical system can be approximated by a sum of K coherent systems, and K is taken as 24 as in other works [21,23,24]. The photoresist model simulates the pattern on the wafer after exposing the photoresist to light, followed by post-exposure baking and development process. In this study, we adopt a simple photoresist model, which has also been utilized in other works [21-25], specifically, the step function model. The expression for this model is as follows:

$$Z(x, y) = \begin{cases} 1, & \text{if } I(x, y) \geq I_{th} \\ 0, & \text{if } I(x, y) < I_{th} \end{cases} \quad (2)$$

were Z presents photoresist pattern on wafer, $I_{th} = 0.225$ is a constant in our implementation.

**B. Level Set Method**

The level set method is a numerical technique for boundary evolution and shape modeling that facilitates the numerical computation of evolving curved surfaces on a Cartesian grid [27]. In two dimensions, the level set algorithm employs the zero-level set of the level set function to represent a closed curve ($\{(x,y)|\psi(x,y) = 0\}$) in the plane, employing the level set function to handle the closed curve. The level set-based ILT method optimizes the masks by performing the evolution of the level set function along the descending direction of the pattern error. The level set of the mask is computed by the expression:

$$\psi(x, y) = \begin{cases} -d(x, y), & \text{if } (x, y) \in \text{inside}(C) \\ 0, & \text{if } (x, y) \in C \\ d(x, y), & \text{if } (x, y) \in \text{outside}(C) \end{cases} \quad (3)$$

where, $\psi(x,y)$ is level set function, and $d(x,y)$ denotes the minimum Euclidean distance from point (x, y) to the mask boundary (C). As mentioned above, we employ the equation to obtain the mask ($M(x,y)$) from the level set in the form of:

$$M(x,y) = \begin{cases} 1, & \text{if } \psi(x,y) \leq 0 \\ 0, & \text{if } \psi(x,y) > 0 \end{cases} \tag{4}$$

The expression for the level set function evolving over time t is as follows:

$$\frac{\partial \psi}{\partial t} = -v|\nabla \psi| \tag{5}$$

where $v$ is the closed curve moving velocity in the normal direction, and $\nabla \psi$ is the first order spatial differentiation of the level set. The above equation is a partial differential equation and its solution is computed iteratively using the finite difference method. Therefore, we use the term $\psi_i(x,y)$ to denote $\psi(x,y,t_i)$. Consequently, the level set at step i+1 ($\psi_{i+1}(x,y)$) is calculated by the expression:

$$\psi_{i+1}(x,y) = \psi_i(x,y) + \Delta t \frac{\partial \psi_i}{\partial t} \tag{6}$$

where $\Delta t$ is the iteration step size, and the evolved closed curve can be obtained after a total of T iterations.

## 3. INVERSE LITHOGRAPHY PHYSICS-INFORMED DEEP NEURAL LEVEL SET METHOD

As illustrated in Fig. 1, the training process of ILDLS is primarily divided into two steps: the pre-training process and the training process which is constrained by inverse lithography physics. The first step, pre-training process, aims to accelerate the second step, which involves re-updating the neuron weights of the model. The second step is further divided into three parts: the pre-trained DL model, the lithography simulation model, and the level set-based ILT correction layer. During forward propagation, the mask predicted by the DL model is simulated by the lithography simulation model to obtain a wafer pattern, and then the pattern error is calculated. In back propagation, the neuron weights of the DL model are corrected using the gradient of the level set-based ILT correction layer by the chain rule. Ultimately, the trained DL model can directly generate masks for unseen test layouts.

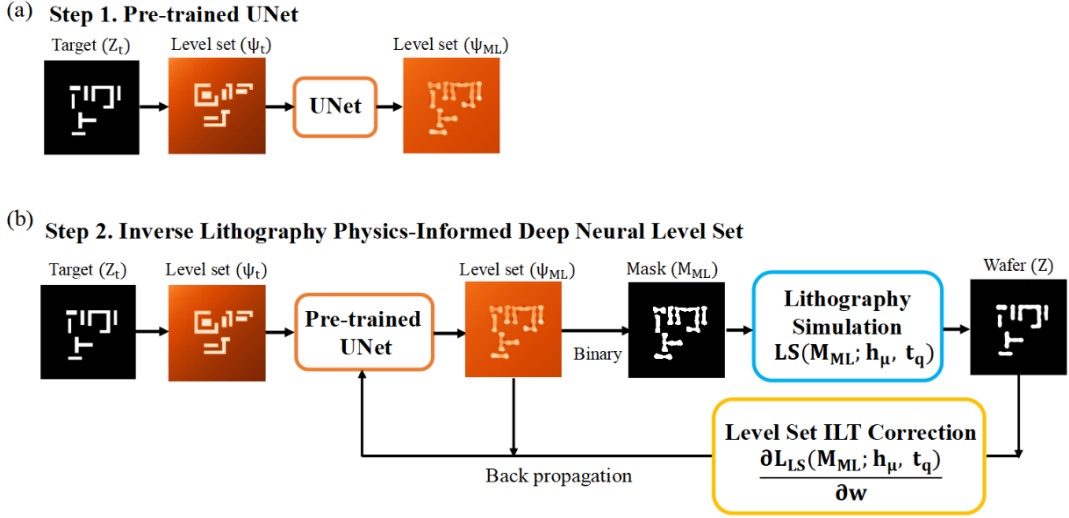

**Fig. 1.** Workflow of the inverse lithography physics-informed deep neural level set method. The training process of the method is divided into two steps. The first step is based on the training dataset obtained by the level set-based inverse lithography technology (ILT), resulting in a pre-trained UNet model. The second step consists of three parts: the pre-trained UNet, the lithography imaging model, and the level set-based ILT correction layer. During forward propagation, the UNet predicts the level set and corresponding mask, and then the fidelity error function is calculated based on the wafer pattern obtained by the lithography imaging model. For back propagation, the neuron weights are updated according to the level set-based ILT correction layer.

**A. Pre-Training Model**

In DL, the pre-training step is crucial in order to generate optimal initial neuron values, because the down streaming loss function optimization method, e.g. gradient descent (GD) method, relies notably on the initial values, thus the model accuracy highly dependent on these initial values. Moreover, if the initial value set is far from the vicinity of the optimal one, it may take the GD method a larger number of iterations to converge. Therefore, a pre-training model can help the following training process regarding model accuracy as well as run time.

Inputting the target layout into the DL model, and then predicting mask, is similar to the image-to-image mapping in computer vision. The most commonly used method in this scenario is the UNet model [28], which is also widely used in pixel-level ILT

[15,23,24]. This model contains a down-sampling path of the input target layout by an encoder consisting of multiple convolutional layers to capture the texture information of the layout. It then performs pixel-level mask prediction using a decoder consisting of multiple deconvolutional layers and skip connections strategy. Theoretically, the convolution operation of convolutional layers is consistent with the convolution in the lithography imaging model, and thus the predicted mask is closer to that of ILT. The training process inputs a series of level set of target layouts ($\psi_t$) and then predicts a series of level set ($\psi_{ML}(\psi_t; w)$). The weights of the pre-trained UNet are obtained by minimizing the following loss function through GD method, calculated as follows:

$$w_{pre} = \mathrm{argmin}_w ||\psi_{ML}(\psi_t; w) - \psi_{ILT}||^2 \quad (7)$$

where w is the neuron weights and $\psi_{ILT}$ are the level sets obtained by level set-based ILT. However, the pattern fidelity of the mask predicted by UNet is not as good as that of the level set-based ILT for the following reasons: firstly, when the geometry of the layout is complex and the pitch is small, the predicted masks exhibit lower printability than level set-based ILT; secondly, the training process of the model does not incorporate any domain knowledge of inverse lithography physics, resulting in poor print fidelity. In summary, it is essential to integrate the physical knowledge of inverse lithography, using the level set-based ILT layer to conduct mask prediction and correction to improve mask printability.

**B. Level Set based ILT Correction Layer**

To achieve the capability of the level set-based ILT layer to conduct mask optimization, it is necessary to integrate it as a layer of the DL into the pre-trained model. The level set-based ILT layer aims to minimize the difference between the wafer and target pattern. The minimized assembly total pattern error $L_{LS}$ is the same as other works [29, 30], which has the following form:

$$L_{LS} = \sum_\mu^U \sum_q^Q \xi(h_\mu) \zeta(t_q) L_{Aerial} \quad (8)$$

where $h_\mu$ and $t_q$ are the defocus and exposure latitude sampling points, respectively, and U and Q represent the numbers of corresponding sampling points, respectively.

$\xi(h_\mu)$ denotes the statistical distribution function of $h_\mu$ with $\xi(h_\mu) = \exp\{-\frac{(h_\mu)^2}{2(\sigma_h)^2}\}$ and $\zeta(t_q)$ represents the statistical distribution function of $t_q$ with $\zeta(t_q) = \exp\{-\frac{(t_q)^2}{2(\sigma_q)^2}\}$, $\sigma_h$ and $\sigma_q$ denotes the standard deviation of $\xi(h_\mu)$ and $\zeta(t_q)$, respectively. These distribution functions are usually obtained by fitting the experimental measurement data. When considering process variations, $h_\mu$ and $t_q$ take values of (-80 nm, 0 nm, 80 nm) and (-0.1, 0.0, 0.1), respectively, and $\sigma_h$ and $\sigma_q$ set to 80 nm and 0.1 in this paper, respectively, resulting in a total of 9 sampling points. If the process variations are not considered, $h_\mu$ and $t_q$ are set to 0 nm and 0%, respectively. $L_{Aerial}$ represents the pattern error to evaluate the difference between the wafer and target pattern under the certain lithography process, in the form of

$$L_{Aerial} = \sum_{x=1}^{N}\sum_{y=1}^{N}(Z(x,y;h_\mu,t_q) - Z_t(x,y))^\gamma \qquad (9)$$

where $Z$ and $Z_t$ are the wafer pattern and target layout, respectively, and $\gamma$ is the tunable parameter and sets to 2. In order to obtain the gradient of the pattern error about mask ($\frac{\partial L_{Aerial}}{\partial M}$), $Z$ as a binary matrix must be continuous to be differentiable. Binary constraints are typically relaxed with a sigmoid function, so the wafer pattern ($Z$) is approximated by a sigmoid function with the following expression:

$$Z(x,y;h_\mu,t_q) = \frac{1}{1+\exp\{-\theta_Z\left[I(x,y;h_\mu) - \frac{I_{th}}{1+t_q}\right]\}} \qquad (10)$$

where $I(x,y;h_\mu)$ represents the aerial image, and $\theta_Z$ is an adjustable parameter that controls the steepness of the sigmoid function and sets to 50. $\frac{\partial L_{Aerial}}{\partial M}$ is represented as follows:

$$\frac{\partial L_{Aerial}}{\partial M} = \gamma(Z - Z_t)^{\gamma-1}\odot\frac{\partial Z}{\partial M} \qquad (11)$$

$$= \gamma\theta_Z\{H^{flip}\otimes[(Z-Z_t)^{\gamma-1}\odot Z\odot(1-Z)\odot(M\otimes H^*)]$$

$$+ (H^{flip})^*\otimes[(Z-Z_t)^{\gamma-1}\odot Z\odot(1-Z)\odot(M\otimes H)]\}$$

where H is a series of optical kernels, $H^*$ is the complex conjugate of H, $H^{flip}$ is the 180° rotation of H, and $\odot$ represents the matrix dot product. Thus, the mask boundary evolution velocity (v) has the following form:

$$v = \sum_{\mu}^{U} \sum_{q}^{Q} \xi(h_{\mu})\zeta(t_q) \frac{\partial L_{Aerial}}{\partial M} \quad (12)$$

Thus, the level set evolution equation according to Eq.6 is expressed as follows:

$$\psi_{i+1}(x,y) = \psi_i(x,y) + \Delta t (\sum_{\mu}^{U} \sum_{q}^{Q} \xi(h_{\mu})\zeta(t_q) \frac{\partial L_{Aerial}}{\partial M})|\nabla \psi_i| \quad (13)$$

Analogous to the GD method, the gradient of $L_{LS}$ about level set ($\frac{\partial L_{LS}}{\partial \psi}$) as follows:

$$\frac{\partial L_{LS}}{\partial \psi} = (\sum_{\mu}^{U} \sum_{q}^{Q} \xi(h_{\mu})\zeta(t_q) \frac{\partial L_{Aerial}}{\partial M})|\nabla \psi| \quad (14)$$

In order to control the smoothness of the mask pattern and to eliminate noise, a penalty function $R_{TV}$ is used in the paper and the expression for this term is given below [31]:

$$R_{TV} = \sqrt{(\frac{\partial \psi}{\partial x})^2 + (\frac{\partial \psi}{\partial y})^2} \quad (15)$$

The penalty term makes the boundary move in the normal direction with a velocity defined by:

$$\kappa = \nabla \cdot R_{TV} = \nabla \cdot (\frac{\nabla \psi}{|\nabla \psi|}) \quad (16)$$

After final consideration of penalty items, the expression for $\frac{\partial L_{LS}}{\partial \psi}$ is as follows:

$$\frac{\partial L_{LS}}{\partial \psi} = [\sum_{\mu}^{U} \sum_{q}^{Q} \xi(h_{\mu})\zeta(t_q) \frac{\partial L_{Aerial}}{\partial M} + \lambda \nabla \cdot (\frac{\nabla \psi}{|\nabla \psi|})]|\nabla \psi| \quad (17)$$

where $\lambda$ is an adjustable parameter, here taken as 0.01.

## C. Inverse Lithography Physics-Informed Deep Neural Level Set Approach

As shown in Fig. 1, the method is divided into three blocks: the first one is to train the pre-trained UNet model based on the level set of the target layout and the optimized level set obtained by the level set-based ILT; the second one is the lithography imaging model utilized to calculate the total fidelity error in forward propagation; and the third one is the level set-based ILT correction layer, which computes the gradient of total fidelity error in backward propagation. The pattern error and its gradient computation mechanism of lithography simulation closely resemble the mechanism of DL forward

and backward propagation. The total loss function for re-training the DL model (Loss) is shown below:

$$\text{Loss} = L_{fit} + \alpha L_{LS} \tag{18}$$

where $L_{fit}$ and $L_{LS}$ are the level set prediction loss function and total pattern error, respectively, and $\alpha$ is the adjustable parameter and sets to 0.008. Consequently, the updated neuron weights of the UNet model are obtained by minimizing the above loss function, which is calculated as follows:

$$w_{opt} = \text{argmin}_w (\overbrace{||\psi_{ML}(\psi_t; w) - \psi_{ILT}||^2}^{L_{fit}} \tag{19}$$

$$+ \alpha \overbrace{\sum_\mu^U \sum_q^Q \xi(h_\mu) \zeta(t_q) L_{Aerial}(M_{ML}; h_\mu, t_q))}^{L_{LS}}$$

where $||\cdot||^2$ stands for the $l_2$ norm. In back propagation, the gradient of total loss function about the neuron weights is obtained by chain rule and Eq.17 as follows:

$$\frac{\partial \text{Loss}}{\partial w} = \frac{\partial L_{fit}}{\partial w} + \alpha \frac{\partial L_{LS}}{\partial w} \tag{20}$$

$$= \frac{\partial L_{fit}}{\partial w} + \alpha \frac{\partial L_{LS}}{\partial \psi_{ML}} \frac{\partial \psi_{ML}}{\partial w}$$

$$= \frac{\partial L_{fit}}{\partial w} + \alpha [\sum_\mu^U \sum_q^Q \xi(h_\mu) \zeta(t_q) \frac{\partial L_{Aerial}(M_{ML}; h_\mu, t_q)}{\partial M_{ML}}$$

$$+ \lambda \nabla \cdot (\frac{\nabla \psi_{ML}}{|\nabla \psi_{ML}|})] |\nabla \psi_{ML}| \frac{\partial \psi_{ML}}{\partial w}$$

where $M_{ML}$ is the mask of $\psi_{ML}$ obtained by Equation 4, and $\frac{\partial L_{fit}}{\partial w}$ and $\frac{\partial \psi_{ML}}{\partial w}$ are numerically computed using automatic differentiation techniques developed by PyTorch, respectively. The method updates the neuron weights based on the above gradient using the Adam optimizer. This approach allows the physical knowledge of inverse lithography physics to be injected into the weights of the UNet model during the training process, resulting in better pattern fidelity. Furthermore, the entire forward and backward propagation process can be seamlessly integrated into unified CUDA-compatible DL toolkits, such as PyTorch or TensorFlow, fully leveraging their computational efficiency. Finally, the updated UNet model can be applied to unseen target layouts to predict the level set and the corresponding mask.

## 4. RESULTS AND DISCUSSION

The ILDLS framework, developed based on PyTorch and CUDA, is developed and tested on a Linux machine with a 2.5 GHz Intel Xeon CPU and a single Nvidia Tesla V100 GPU. The target layout datasets are obtained from the two works of GAN-OPC and Neural-ILT, which synthesize layouts based on the design specifications from existing 32 nm M1 layout topologies with minimum critical dimension of 80 nm [21,23]. The level set-based ILT lithography engine used to generate the masks is the recently published GPU-enabled level set-based ILT optimizer [32]. Specifically, the CGD method combined with the gradient of Eq. 17, is used to optimize the level sets to obtain the corresponding masks. Additionally, its core computational functions are implemented using CUDA, ensuring high computational efficiency.

Table 1. Comparison of the average edge distance error (AEDE) and the difference between maximum and minimum (|Max-Min|) values of EDE of level set-based ILT (ILT), UNet, ILDLS, UNet combined with a few additional iterations of level set-based ILT (UNet+ILT), and ILDLS+ILT methods. The values in bold represent outperformance on the same task.

| Method | AEDE (nm) | |Max-Min| (nm) |
|---|---|---|
| ILT | 5.67 | 28.20 |
| UNet | 7.79 | 27.28 |
| ILDLS | 5.89 | 13.00 |
| UNet+ILT | 4.91 | 14.99 |
| ILDLS+ILT | **4.02** | **11.83** |

### A. The effectiveness of LPDLS

In this section, we aim to validate the effectiveness of the proposed approach by dividing the dataset into a training dataset (data ratio is 80%) and a test dataset (data ratio is 20%). Both the pre-trained model and the LPINDS are trained on the training dataset, and the comparison of the results for the different methods is based on the test dataset. The level set-based ILT is taken as a baseline method.

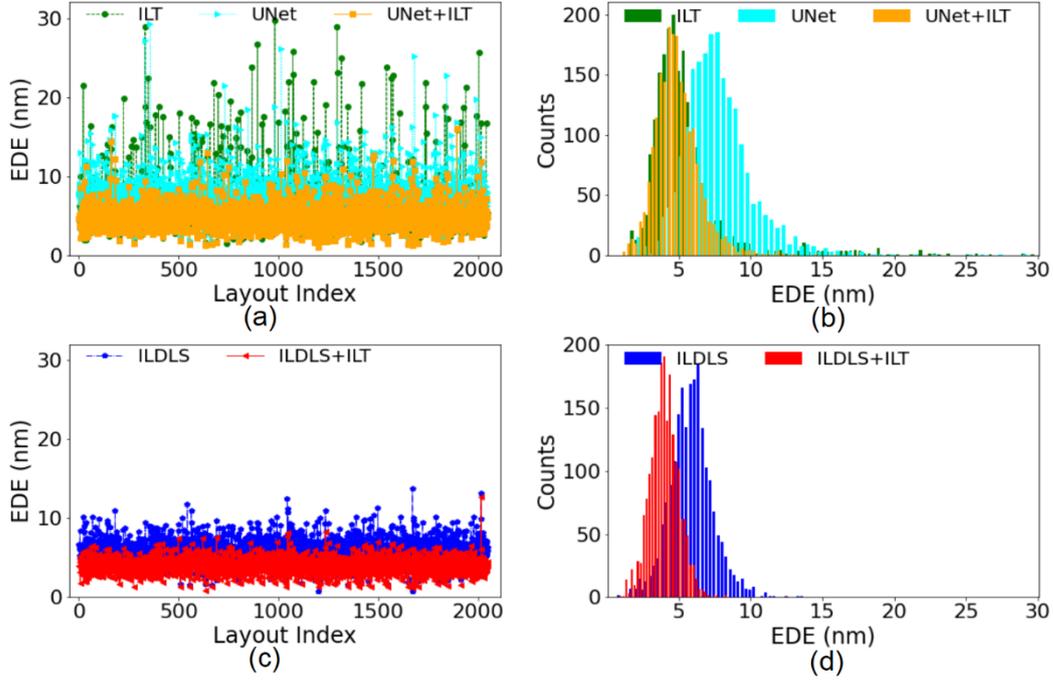

**Fig. 2.** Comparison of edge distance error (EDE) of masks obtained by level set-based ILT (ILT), UNet, UNet combined with a few additional iterations of level set-based ILT (UNet+ILT), ILDLS, and ILDLS+ILT methods on the test dataset. (a) and (c) EDE of each mask. (b) and (d) Histograms of the distribution of EDE.

Figure 2 illustrates the pattern fidelity of the level set-based ILT, UNet, UNet combined with a few additional iterations of level set-based ILT (UNet+ILT), ILDLS, and ILDLS+ILT on the test dataset. Here, the pattern fidelity is further improved by a few additional iterations (as a common strategy nowadays) of level set ILT which applies finer corrections to the predicted mask. We adopt the edge distance error (EDE) as the evaluation metric of printability, which has been adopted by previous works due to its insensitivity to mesh size, and is defined as $EDE = \frac{A_{shadow}}{L}$, where $A_{shadow}$ and L represent the absolute difference in area between the wafer and the target pattern and the sum of the perimeter of target patterns, respectively, which can be understood as the weighted average of the edge placement error [33,34]. As shown in Fig. 2(a) and (c), compared with the level set-based ILT, ILDLS method generally reduces the EDE of almost all the masks substantially. However, the level set-based ILT obtains a larger

EDE for some of the masks, and thus the robustness of our approach is better than that of the level set-based ILT. In addition, it is found that the EDE of mask predicted by ILDLS+ILT can be further reduced universally. Fig. 2(b) and (d) show the EDE distributions of the masks generated by different methods. We find that the most of EDEs of the level set-based ILT are distributed between 1.5-10 nm and have a long-tailed distribution up to 30 nm. However, the vast majority of results for ILDLS are distributed in the range of 0.6-10 nm without long-tailed distribution. The most of the ILDLS+ILT results are distributed in the range of 0.8-7 nm and there is also no long-tail distribution. Once again, ILDLS, especially ILDLS+ILT method, shows better results in comparison with the level set-based ILT due to the fact that ILDLS incorporated the level set-based ILT layer. Comparison details are displayed in Table 1, where it could be seen that the average EDE of ILDLS is 5.89 nm, which is close to the level set-based ILT result of 5.67 nm, whereas the ILDLS+ILT result is 4.02 nm, which has a 29.1% reduction in EDE compared to level set ILT, indicating that both ILDLS and ILDLS+ILT could generally reduce the fidelity error. In addition, we use the difference between the maximum and minimum values of the EDE (|Max-Min|) to evaluate the robustness of the method, and the results show that the |Max-Min| of ILDLS and ILDLS+ILT are 13.00 and 11.83 nm, respectively, which are reduced by 53.9% and 58.0% compared to 28.20 nm of level set ILT, which proves that the stronger robustness of our method. The computation time of the algorithm is crucial for its application. Fig. 6(a) shows that the average computational time for ILDLS to predict a single mask of clip is a few hundredths of that of level set ILT, and the computation time for ILDLS+ILT is about one-tenth of that of ILT, so the method can significantly accelerate ILT.

Figures 3 and 4 show the masks predicted by different methods for two optional layouts from the test dataset, corresponding to the wafer pattern and the absolute difference between the wafer and the target pattern. Figures 3(a)-(d) and 4(a)-(d) clearly show the missing parts of the pattern on the wafer compared to the target layout in the results of ILT. However, Figs. 3(i)-(l) and 4(i)-(l) present the results of ILDLS, where the pattern on the wafer is intact compared to the level set-based ILT. As shown in

Figures 3(m)-(p) and 4(m)-(p), ILDLS+ILT further corrects the mask to improve pattern fidelity. Among the masks generated by the above three methods shown in Fig. 3, it is evident that the degree of mask distortion in the missing parts obtained by the latter two methods is higher than that of the level set-based ILT. The above results demonstrate the effectiveness of ILDLS and our approach demonstrates advantages over level set-based ILT. It is because the level set ILT employs a GD algorithm, therefore the search space is relatively small. However, the powerful high-dimensional data fitting capability of DL ensures that the search space is much larger than that of ILT. In particular, the level-set ILT correction layer integrated into the DL guarantees that the mask optimization with constraints of inverse lithography physics.

### B. The necessity of LPDLS

To show the necessity of our method, we evaluate our method and the UNet method on the test dataset. Fig. 2(a) and (c) show the EDEs of the UNet, UNet+ILT, ILDLS, and ILDLS+ILT methods. It is found that some masks generated by the UNet method exhibit high EDEs, while the reduction in EDE is still limited using UNet+ILT. However, the ILDLS and ILDLS+ILT methods show substantial reductions in EDEs for almost all layouts, which proves that the integration of level-set ILT layers into the DL can significantly improve the pattern fidelity. As shown in Fig. 2(b) and (d), most of the EDEs of UNet are distributed in the range of 2-15 nm with a long-tailed distribution up to 30 nm. Although UNet+ILT could reduce the EDEs overall, the vast majority of the EDEs are distributed in the range of 1-10 nm, which is still large. In contrast, the most of EDEs of ILDLS and ILDLS+ILT are in the range of 0.6-10 nm and 0.8-7 nm, respectively. Therefore, the range of the EDEs is compressed, demonstrating the higher suitability of our methods. The comparative quantitative results are shown in Table 1, where it can be seen that the average EDE of ILDLS is 24.4% lower than that of UNet, while the result of ILDLS+ILT is 18.1% lower than that of UNet+ILT. In addition, the |Max-Min| of ILDLS and ILDLS+ILT are 52.3% and 21.1% lower than that of UNet and UNet+ILT, respectively. As shown in Fig. 6(a), we find

that the average computation time for a single mask predicted by ILDLS is almost the same as that of the UNet method. Therefore, our method can generate masks with higher pattern quality in the same run time level, with stronger printability robustness.

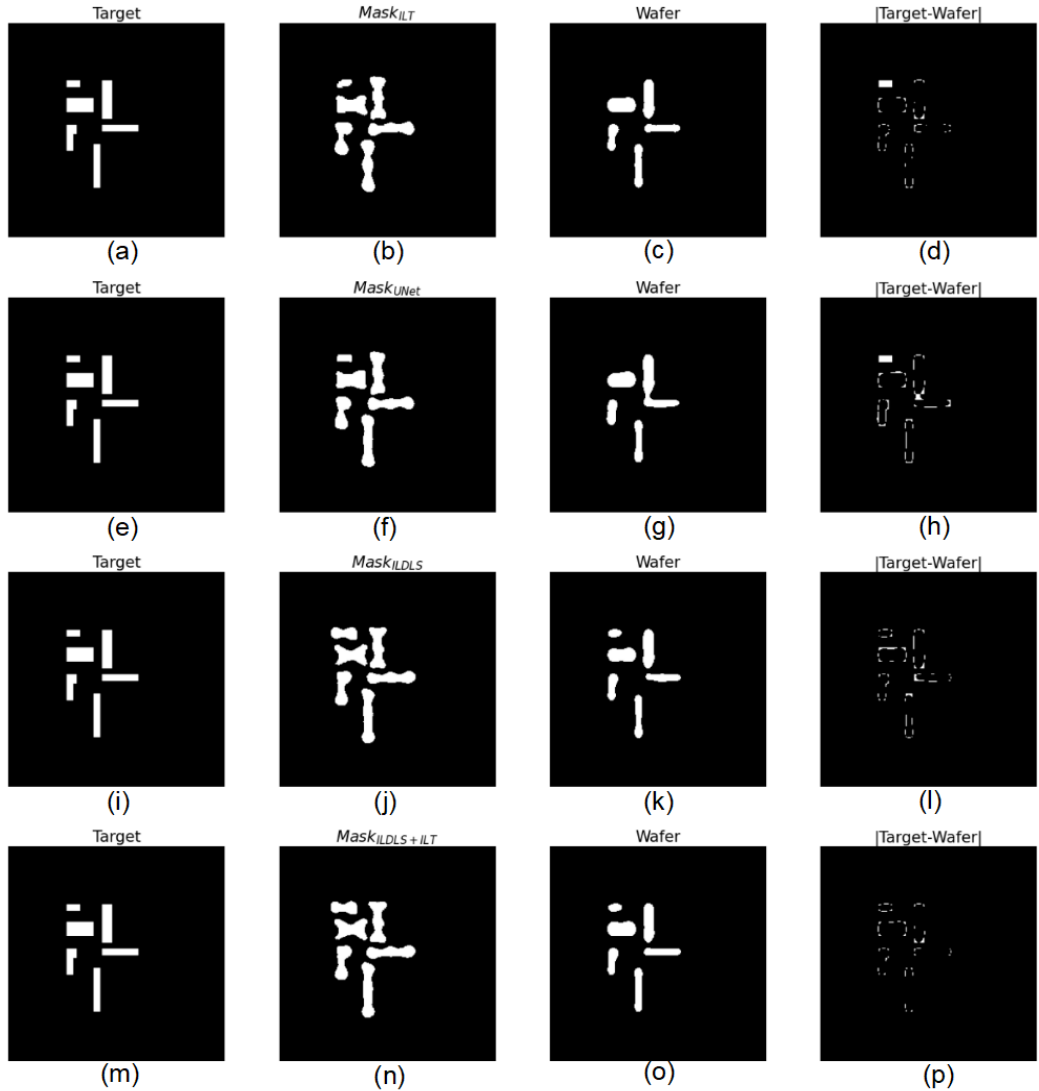

**Fig. 3.** Comparison of the masks obtained by the level set-based ILT, UNet, ILDLS, and ILDLS+ILT methods on an optional target layout from the test dataset, corresponding to the wafer pattern and the absolute difference between the wafer pattern and the target pattern. Results of the (a)-(d) ILT, (e)-(h) UNet, (i)-(l) ILDLS, and (m)-(p) ILDLS+ILT.

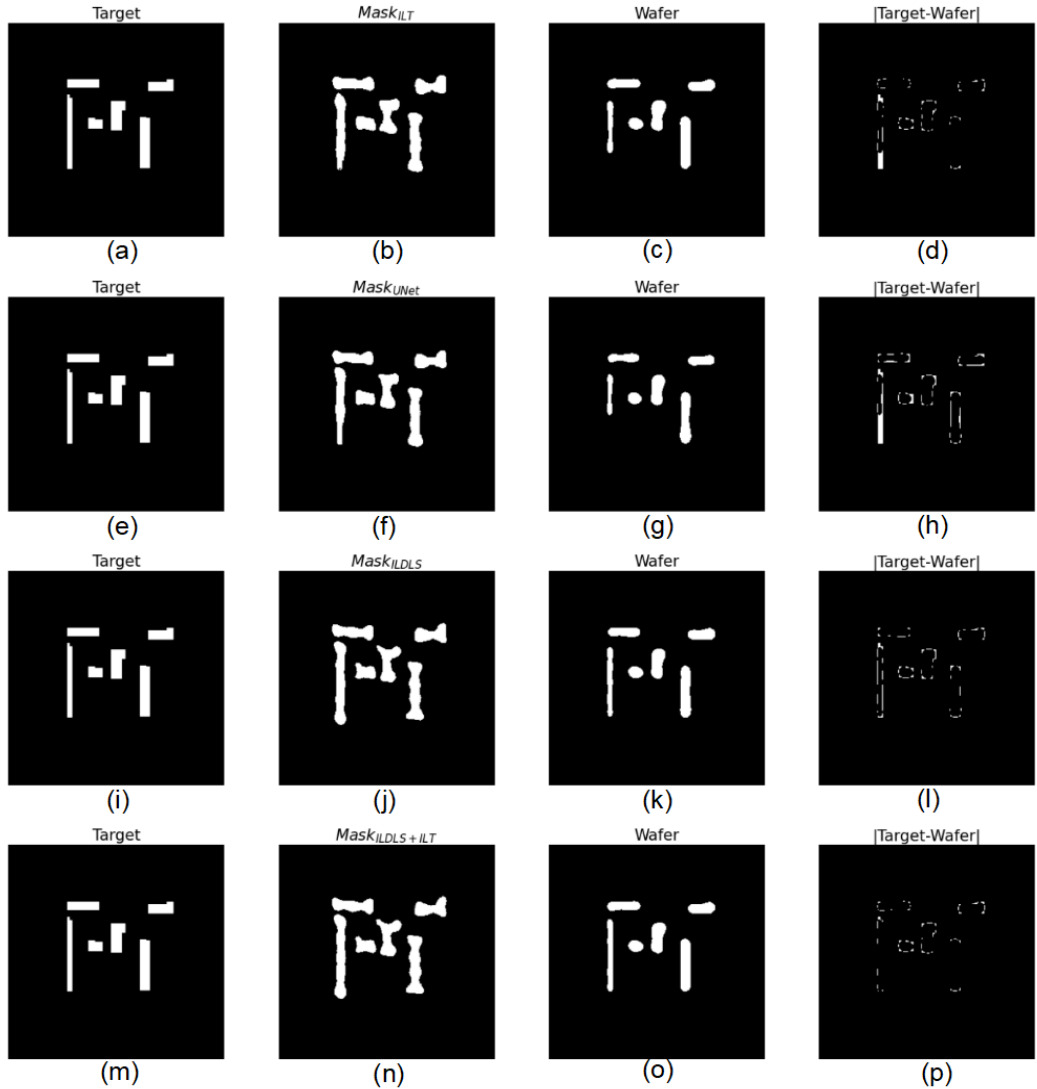

**Fig. 4.** Comparison of the masks obtained by the level set-based ILT, UNet, ILDLS, and ILDLS+ILT methods on an optional target layout from the test dataset, corresponding to the wafer pattern and the absolute difference between the wafer pattern and the target pattern. Results of the (a)-(d) ILT, (e)-(h) UNet, (i)-(l) ILDLS, and (m)-(p) ILDLS+ILT.

Figures 3(e)-(h) and 4(e)-(h) clearly show that the UNet method does not have any improvement regarding the missing patterns on the wafers obtained by level set-based ILT and even generates a hotspot of hard bridge (Fig. 3(g)). The reason is that the training process of the UNet model simply tries to replicate masks obtained by the level set-based ILT, and the predicted errors further degrade the pattern quality. However,

Figures 3(i)-(l) and 4(i)-(l) demonstrate that the pattern fidelity of the mask predicted by ILDLS is significantly better than the UNet, with a more complete wafer pattern compared to the latter and the elimination of bridging hotspot. Comparison with the UNet results proves that ILDLS is essential, especially their computation times of prediction are almost the same. It's mainly because although DL is very powerful in high-dimensional function approximation, the search space is too large to generate masks with optimal fidelity. However, domain knowledge of inverse lithography physics can help limit the search space and effectively guide the searching process, so that the mask optimization process can be more efficient.

### C. ILDLS with Lithography Process Variations

In order to verify the effectiveness and necessity of the proposed method considering the variation of the lithography process, we optimize the level set functions and their corresponding masks using the level set-based ILT with the lithography process variations (ILT-PV), and therefore the results of the ILT-PV serve as a baseline.

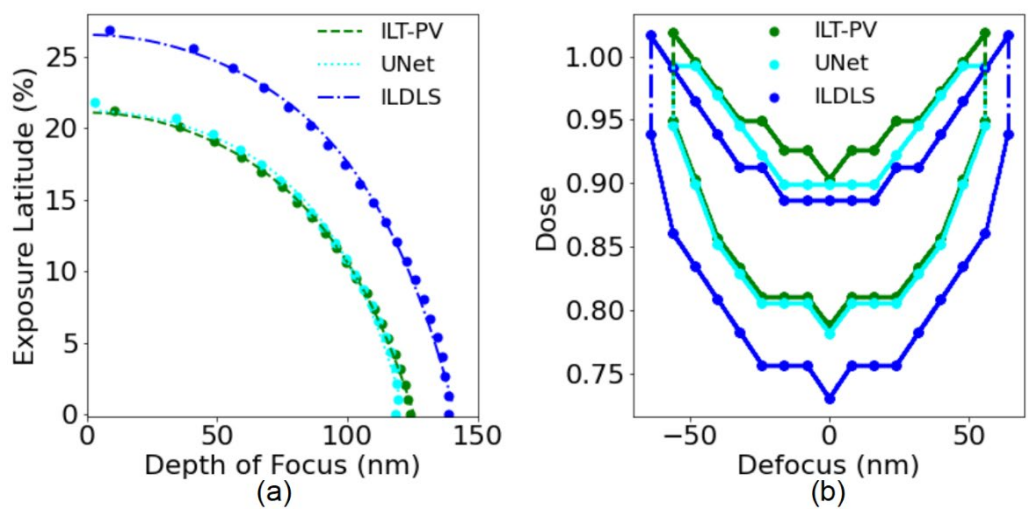

**Fig. 5.** Comparison of the process windows of the masks obtained by the level set-based ILT with lithography process variations (ILT-PV), UNet, and ILDLS methods on the test dataset. (a) Curve of exposure latitude versus depth of focus. (b) Exposure dose versus defocus.

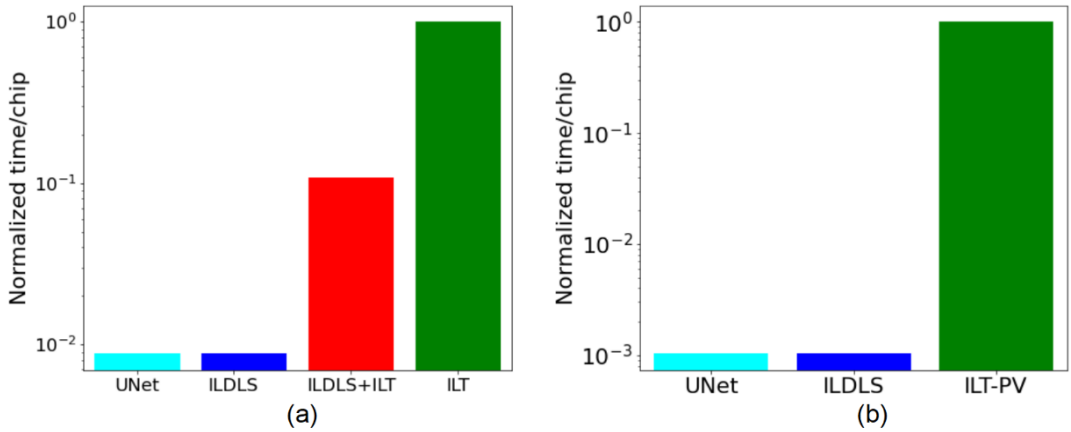

**Fig. 6.** (a) Comparison of average computation time for a single mask generated by level-set ILT, UNet, ILDLS, and ILDLS+ILT approaches. (b) Comparison of average computational time consumption for a single mask obtained by ILT with lithography process variations (ILT-PV), UNet, and ILDLS methods.

**Table 2. Process window (PW) and worst image log slope (Worst ILS) of masks obtained by the level set-based ILT lithography process variations (ILT-PV), UNet, and ILDLS methods on test dataset. PW was evaluated using the exposure latitude (EL) upper bound at a depth of focus (DOF) of 0 nm (EL@DOF=0), and the DOF upper bound at 5% EL (DOF@%EL). The values in bold represent outperformance on the same task.**

| Method | DOF@5%EL | EL@DOF=0 | Worst ILS |
|--------|----------|----------|-----------|
| ILT-PV | 116.0    | 21.1     | 39.1      |
| UNet   | 113.9    | 22.0     | 42.0      |
| ILDLS  | **133.9**| **26.5** | **49.6**  |

Fig. 5(a) illustrates the results of exposure latitude (EL) versus depth of focus (DOF) from different methods, i.e., the maximum EL allowed for a certain DOF, and the maximum DOF allowed for a certain EL to ensure pattern fidelity. The larger the enclosure area of the curve indicates the less sensitive the mask is regarding lithography process variations, implying better manufacturing yield. As shown in Fig. 5(a), ILDLS method could enhance the PW compared to UNet and ILT-PV. The detailed results in

Table 2 show that when the range of EL is within 5%, the DOF of ILDLS-generated masks has to be no more than 133.9 nm, which is 15.4% and 17.6% higher than that of ILT-PV and UNet, respectively. The EL upper bound at 0 nm DOF for masks predicted by ILDLS is 26.5%, which is improved by 25.6% and 20.5% over ILT-PV and UNet, respectively. Fig. 5(b) illustrates the PW of exposure dose versus defocus for masks generated by different methods. It could be seen that the PW from the ILDLS method is larger than those from the other two methods, where larger PW indicates more stable yield that a mask could have. In addition, the image log slope (ILS) evaluates the quality of the aerial image, where the larger value implies higher aerial imaging contrast. As shown in Table 2, the worst ILS of the mask predicted by ILDLS is improved by 26.9% and 18.1% compared to ILT-PV and UNet, respectively. As shown in Fig. 6(b), the average computation time of a single mask predicted by ILDLS is almost identical to that of UNet, and the computational efficiency is improved by hundreds of times compared with ILT-PV.

## 5. CONCLUSION

In this paper, we propose an inverse lithography physics-informed deep neural level set algorithm for mask optimization. Firstly, the pre-trained UNet model is obtained based on the level set of the target layout and the optimal level set calculated by the level set-based ILT. Secondly, the level set-based ILT layer is integrated into the DL architecture. During forward propagation, the lithography imaging model is utilized to obtain the wafer pattern and the pattern error. In backward propagation, the gradients of pattern error computed by the level set-based ILT layer are used to correct the neuron weights via the chain rule. Thus, the domain knowledge of inverse lithography is effectively injected into the re-trained model. By applying this method to unseen target layouts to predict masks, we show that the pattern fidelity and PW could be obviously improved compared to DL and ILT, achieving two to three orders of magnitude speedup for ILT computation.